\documentclass[journal]{IEEEtran}
\usepackage[utf8]{inputenc}
\usepackage[T1]{fontenc}
\usepackage[brazil,english]{babel}
\usepackage{graphicx}
\usepackage{amssymb}
\usepackage{graphics}
\usepackage{bibentry}
\usepackage{epsfig}
\usepackage{booktabs}
\usepackage{lscape}
\usepackage{enumitem}
\usepackage{float}
\usepackage{multirow}
\usepackage{amsmath}
\usepackage{textcomp}
\usepackage{adjustbox}
\usepackage{breqn}
\usepackage{cite}
\usepackage{multirow}
\usepackage[usenames, dvipsnames]{color}
\usepackage{makecell}
\usepackage[ruled,vlined]{algorithm2e}
\usepackage{algpseudocode}
\usepackage{tabu}
\usepackage{xcolor}
\usepackage{colortbl}
\usepackage[normalem]{ulem}
\useunder{\uline}{\ul}{}

\ifCLASSINFOpdf
\else
\fi
\begin{document}

\title{Multi-modal Medical Neurological Image Fusion using Wavelet Pooled Edge Preserving Autoencoder }

\author{Manisha Das,~\IEEEmembership{Student Member,~IEEE}, Deep Gupta, ~\IEEEmembership{Senior Member,~IEEE,} Petia  Radeva, ~\IEEEmembership{Fellow,~IAPR,} and Ashwini M Bakde
\thanks{Manisha Das is with the Department
of Electronics and Comm. Engineering, Visvesvaraya National Institute of Technology Nagpur 40010, India. e-mail: (das.manisha1989@gmail.com ).}
\thanks{Deep Gupta is with the Department
of Electronics and Comm. Engineering, Visvesvaraya National Institute of Technology Nagpur, 440010, India. e-mail: (deepgupta@ece.vnit.ac.in).}
\thanks{Petia Radeva is with the Departament de Mathematics and Informatics, Universitat de Barcelona, 08007 Barcelona, Spain, and also with the Computer Vision Center, 08193 Cerdanyola, Spain. e-mail: (petia.ivanova@ub.edu).}
\thanks{Ashwini Bakde is with the Department of Radio-Diagnosis, All India Institute of Medical Sciences Nagpur, 441108, India. e-mail: (ashwini@aiimsnagpur.edu.in).}

}

\maketitle

\maketitle

\begin{abstract}
Medical image fusion integrates the complementary diagnostic information of the source image modalities for improved visualization and analysis of underlying anomalies. Recently, deep learning-based models have excelled the conventional fusion methods by executing feature extraction, feature selection, and feature fusion tasks, simultaneously. However, most of the existing convolutional neural network (CNN)  architectures use conventional pooling or strided convolutional strategies to downsample the feature maps. It causes the blurring or loss of important diagnostic information and edge details available in the source images and dilutes the efficacy of the feature extraction process. Therefore, this paper presents an end-to-end unsupervised fusion model for multimodal medical images based on an edge-preserving dense autoencoder network. In the proposed model, feature extraction is improved by using wavelet decomposition-based attention pooling of feature maps. This helps in preserving the fine edge detail information present in both the source images and enhances the visual perception of fused images. Further, the proposed model is trained on a variety of medical image pairs which helps in capturing the intensity distributions of the source images and preserves the diagnostic information effectively. Substantial experiments are conducted which demonstrate that the proposed method provides improved visual and quantitative results as compared to the other state-of-the-art fusion methods.
\end{abstract}

\section{Introduction}
Medical image fusion has become an integral part of clinical diagnosis due to its powerful ability to provide more conclusive and comprehensive diagnostic interpretations \cite{li2017pixel}. It also plays a very important role in the detection, evaluation, and treatment of various neurological disorders. Prevalent modalities used for neuroimaging include Computed tomography (CT),  magnetic resonance (MR) imaging,  single photon emission computed tomography (SPECT), positron emission tomography (PET), etc. Each of these modalities unmasks some unique information about the anatomy or physiology of the tissues being scanned. Medical image fusion provides a solution to combine the complementary information because a composite visualization of the complementary information of these modalities helps clinicians to draw more conclusive and accurate diagnostic outcomes.

Multi-modal image fusion is carried out in three steps namely feature extraction, feature selection, and feature fusion  \cite{HERMESSI2021108036}. 
For feature extraction, most of the conventional fusion methods use spatial and spectral decomposition, sparse representation, bio-inspired spiking neural networks, fuzzy logic, guided filters, etc \cite{li2017pixel}. However, the fusion accuracy of these methods is dependent on the selection of several factors such as the number of decomposition levels, type of filters, number of dictionary entries, model hyper-parameters, etc. For feature selection, local activities of pixels or transform coefficients based on Laplacian energy, spatial frequency, phase congruency, standard deviation, morphological gradient, etc., are used to calculate the feature sparsity of the source images \cite{dogra2017multi}. The choice of these activity measures is crucial for accurate apprehension of the type of information carried by the pixels or transform coefficients of the source images. The feature fusion step generally uses fusion rules such as choose-max, average, weighted average, etc.  which may dilute the effectiveness of the feature extraction and selection blocks \cite{du2016overview}. For example, a choose-max rule may result in spatial discontinuities with abrupt changes in pixel intensities, resulting in artifacts in a fused image. Similarly, averaging may introduce sudden intensity drops and partial blurring of edges resulting in inferior visual results.

\begin{figure*}[!t]
	\centering
	\scalebox{0.8}{\includegraphics[width=7.4in, 
		]{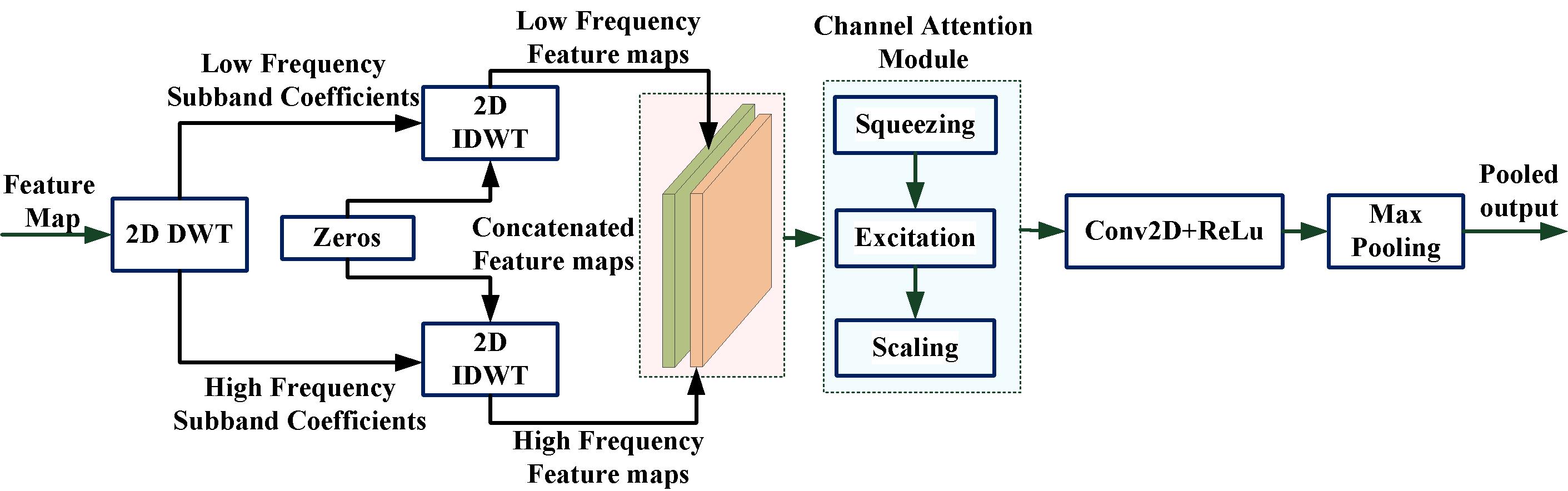}}\\
	\caption{Framework of wavelet decomposition based edge preserving pooling }\label{pooling}
\end{figure*}
In context with the above discussion, it can be stated that the handcrafted feature extraction, selection, and fusion techniques may suffer from various drawbacks and pose a bottleneck in achieving superior fusion performance. Therefore, in this paper, a novel end-to-end medical image fusion method with an edge-preserving U-Net encoder-decoder network is presented in which the max-pooling layer in each convolutional block of the encoder is replaced by  Wavelet decomposition-based edge-preserving pooling (WDEPP) layer. The source images are concatenated and fed to the encoder to extract the edge-preserved feature maps. The decoder block uses the encoded features and reconstructs the fused image. The loss function is also made content adaptive taking into account the preservation of pixel intensity, gradients, and multi-scale structural similarity of the source images being fused. The salient contributions are listed as follows:
\begin{enumerate}
    \item  To the best of our knowledge, it is the first time to utilize edge preserving pooling-based CNN in the area of image fusion.
    \item  An end-to-end medical image fusion method is proposed using a U-Net autoencoder structure with wavelet-based attention pooling layers in the encoder block which has  two key advantages;
    \begin{enumerate}
    \item The feature map is decomposed into approximate and detailed components using wavelet transform before calculating individual channel attention which helps in the effective preservation of both the global contrast and local gradients of the feature maps.
    \item The edge-preserved pooling operation enhances the feature extraction process without losing the sharp edges present in feature maps.
    \end{enumerate}
\end{enumerate}
The rest of the paper is organized as follows. Section 2 discusses the related work and the details of the WDEPP. Section 3 describes the proposed method in detail. In Section 4, the experimental details and result validations are presented, followed by the conclusion in section 5.

\section{Related Work}
\subsection{Deep learning based fusion methods}   
In recent years, deep learning-based fusion methods have escalated in the area of image fusion with their powerful ability to accomplish end-to-end fusion tasks by integrating the three basic fusion steps namely feature extraction, selection, and fusion. Some approaches use convolutional neural networks (CNN) \cite{liu2017medical}, convolutional sparse representation (CSR) \cite{liu2016image}, autoencoders (AE) based supervised models \cite{li2018densefuse} to extract meaningful features from the source images. One major drawback with these methods is that the DL model is used only as a feature extractor, however, the feature selection and fusion are done by conventional handcrafted rules. Some unified approaches with content-adaptive loss functions have also been presented \cite{Xu2022}, however, most of the models are trained on the dataset of other fusion tasks such as visible-infrared fusion, multi-focus fusion, etc., and hence testing on medical image pairs generates poor visual results \cite{zhang2021sdnet, Xu2022}. As no reference is available in the case of medical image fusion, still some approaches with unsupervised learning models such as generative adversarial networks (GANs) \cite{ma2020ddcgan, FU2021} and dense CNN networks \cite{xu2021emfusion, fan1906semantic} are presented and trained using medical images with content oriented loss functions to achieve end-to-end fusion and also provide more convincing fusion results than other models. However, most of these methods concentrate more on feature selection and fusion parts intending to preserve the information available in the source image. However, the feature extraction part is not refined and a little exploration is done in the network architectures. Recently, detail and edge-preserving CNNs have shown better performance in the area of image classification, segmentation, reconstruction, etc \cite{sineesh2021exploring}. These CNNs use feature-preserving pooling with channel attention that helps in preserving the edges in the feature maps. In this paper, we explore the applicability of edge-preserving pooling-based dense networks in the area of medical image fusion and further validate its effect on fusion performance.
\begin{figure*}[!t]
	\centering
	\scalebox{0.70}{\includegraphics[width=7.4in, 
		]{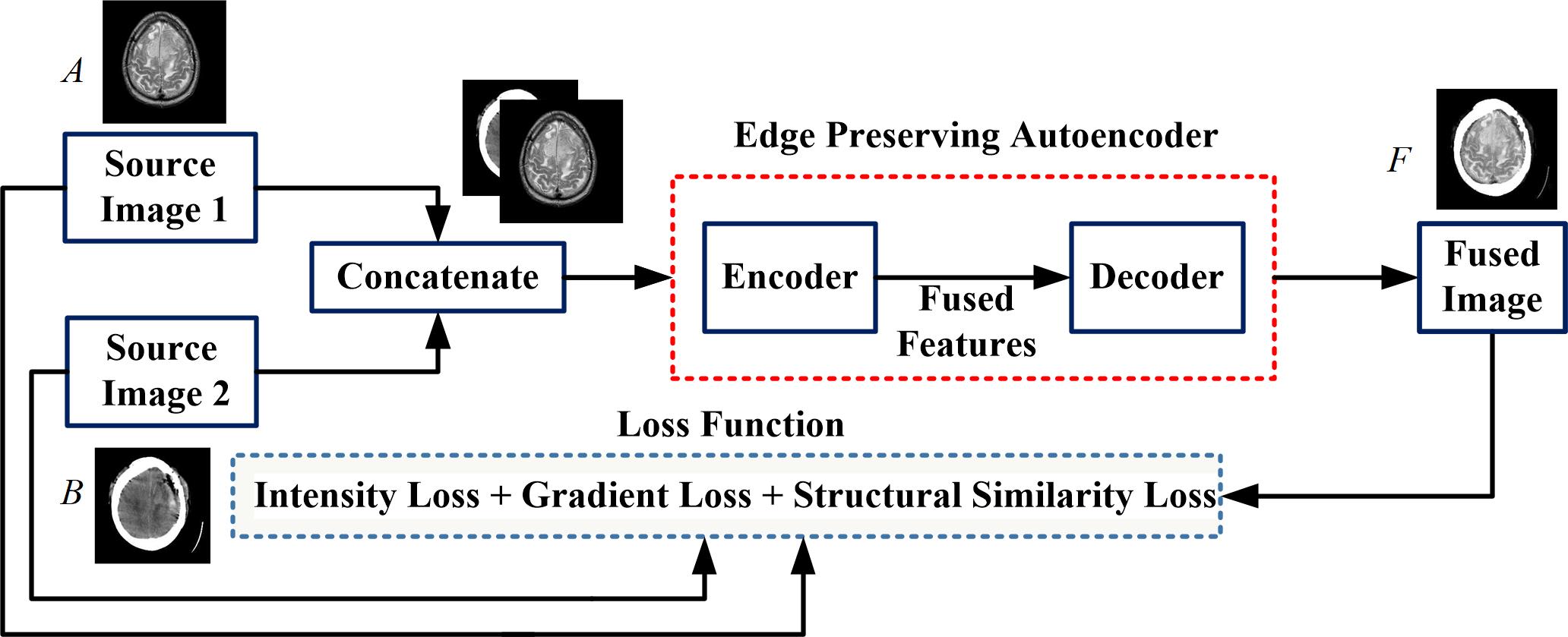}}\\
	\caption{Process flow of the proposed fusion method }\label{bd}
\end{figure*}

\subsection{Wavelet decomposition-based edge preserving pooling}

Wavelet-based pooling techniques have been employed for downsampling in convolutional neural networks (CNNs) to improve robustness against noise as in the case of conventional strided convolution and max-pooling \cite{li2020wavelet}. However, in most of these approaches, the high-frequency components of the feature maps are not considered resulting in the blurring of the down-sampled feature maps. In the wavelet-based edge preserving pooling approach, both the low and high-frequency subbands are utilized for channel attention which provides effective noise-robustness and preserves the edges of the feature maps as well\cite{sineesh2021exploring}.  Fig. \ref{pooling} shows a framework of the wavelet-decomposition-based edge preserving pooling (WDEPP) approach. At first,  the feature map ($X$) of size $H\times W\times C$ is decomposed into low-frequency $LF(X)$ and high-frequency bands $HF(X)$  using single level 2-dimensional discrete wavelet transforms (DWT) as shown in Eq. 1. Here, $H$  and $W$ represent the spatial dimension of the feature maps and $C$ represents the number of channels.
\begin{equation}
\begin{aligned}
\scriptsize
  \left \{ LF(X), HF(X) \right \} = DWT(X)
\end{aligned}
\end{equation}
Further, the individual low and high-frequency bands are concatenated with zeros and subjected to 2D-inverse DWT for individual sub-band reconstruction of the feature maps, and the concatenated feature set $F(X)$  of size $H\times W \times 2C$ is obtained.

\begin{equation}
\begin{aligned}
\scriptsize
F(X) =\left \{ IDWT(zeros,LF(X) ),IDWT(zeros,HF(X)) \right \}
\end{aligned}
\end{equation}
The feature set is then passed through a channel attention module consisting of a squeeze and excitation network ($SqEx$) followed by feature scaling. This module embeds, estimates, and scales the features with attention weights $F_{AW}$ and generates attention-weighted features as the output.
\begin{equation}
\begin{aligned}
\scriptsize
 SqEx\left \{ F(X) \right \}=F_{AW}\times F(X)
\end{aligned}
\end{equation}
Finally, a convolutional layer and rectified linear activation unit (ReLU) are applied to the weighted feature maps for size consistency of the weighted feature maps with the original feature map size i.e. $H\times W \times C$.  At last, the pooled output of size $H/2 \times W/2 \times C$ is obtained by performing a max-pooling operation on the weighted feature maps. 

\section{Proposed Fusion Method}
This section gives the details of the proposed edge preserving autoencoder-based fusion method. We first discuss the overview of the proposed framework followed by detailed discussions on the edge-preserving network architecture and loss functions. 

\subsection{Overview}
The block diagram of the proposed method is shown in Fig. \ref{bd}.  Let $A$ and $B$ represent the source MR image and CT image respectively. These source images are concatenated along the channel dimension and given to the edge-preserving autoencoder. The encoder ($E$) extracts various features of the source images at different scales and orientations. The fused features are then fed to the decoder $(D)$ which reconstructs the fused image. During the training process, the loss function of the network helps to retain the complementary information of both the source images.

During the testing phase, the concatenated source images are given to the trained autoencoder and the fused images are obtained in an end-to-end manner. In the case of MR and SPECT/PET image fusion, the color images are first subjected to $RGB$ to $YUV$ color space conversion \cite{das2022tim, das2020nsst}. The luminescence ($Y$) channel is considered further for fusion as it captures the metabolic activities of the underlying tissue structures. The $Y$ channel is fused with the MR image to generate a single-channel fused image which is then combined with the original $U$ and $V$ components followed by $YUV$ to $RGB$ color space conversion to get the final fused color image.

\subsection{Loss function}
During the training process, the autoencoder is made to learn for preserving the complementary diagnostic information of both the source imaging modalities. Hence, the loss function of the autoencoder is framed accordingly. The total loss ($L_{total}$) consists of three parts namely intensity loss ($L_{intensities}$) , gradient loss ($L_{gradient}$) and structure loss ($L_{structure}$)   corresponding to ensure the retention of intensity distribution,  fine edge details and structural details of variety of tissues,  respectively.  
\begin{equation}
    L_{total}=L_{intensity}+L_{gradient}+L_{structure}
\end{equation}

\begin{figure*}[!t]
	\centering
	\scalebox{0.7}{\includegraphics[width=7.4in, 
		]{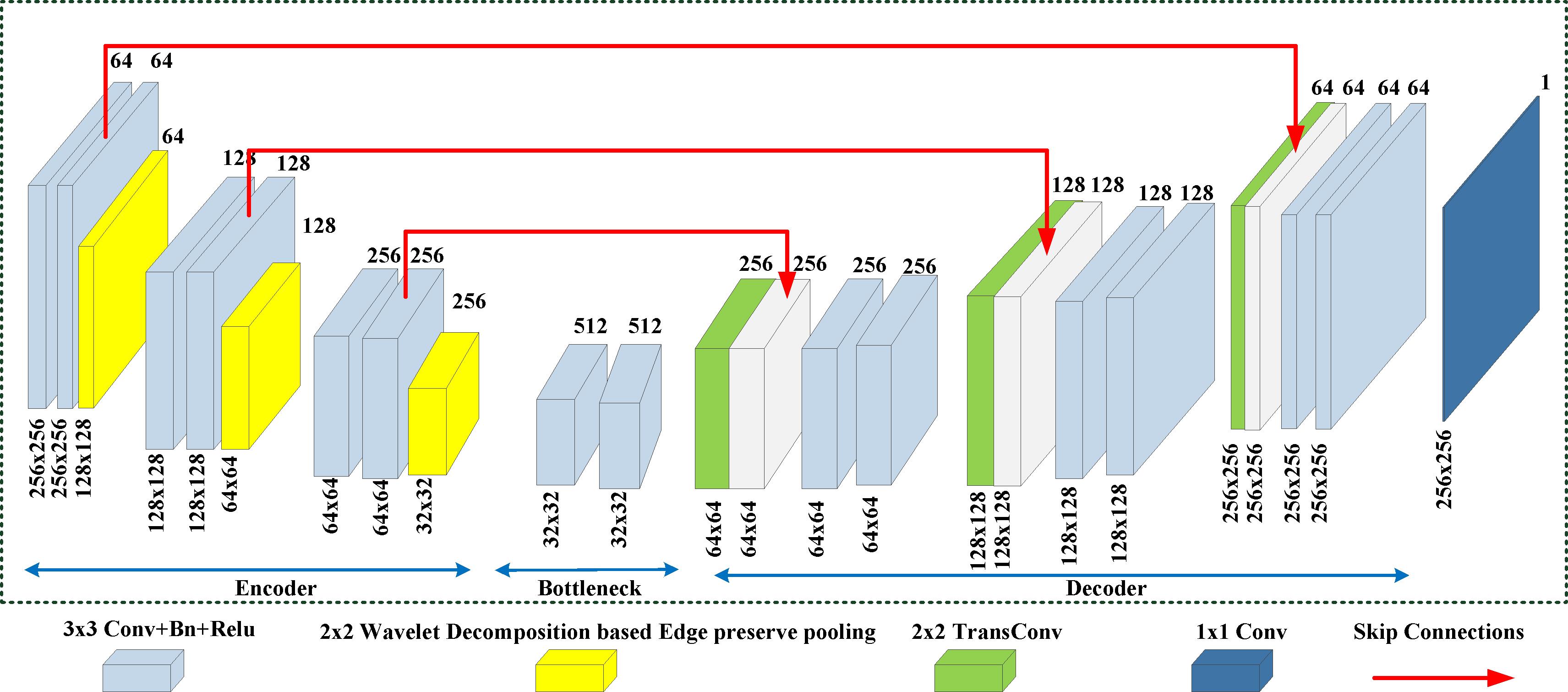}}\\
	\caption{Network architecture }\label{gen}
\end{figure*}

The pixel brightness of the various modalities conveys specific information regarding the underlying tissue structures. In anatomical images, the pixel intensity corresponds to the density of the tissues, for example, in the case of CT image, the dense regions such as bones, and calcification regions appear much brighter as compared to the rest of the less dense tissues. Similarly, the abnormal regions with lesions have much larger intensities as compared to surrounding soft tissues in MR images. For functional images, the higher-intensity regions may indicate abnormal metabolic activity. The intensity of both the source images characterizes unique information that needs to be preserved well in the fused image for accurate diagnosis hence the intensity loss is defined as,
\begin{equation}
\begin{aligned}
\scriptsize
    L_{intensity}=\frac{1}{W\times H}\sum_{i=1}^{W}\sum_{j=1}^{H} (F_{i,j}- max(A_{i,j},B_{i,j}))^{2}
\end{aligned}
\end{equation}
where $W$ and $H$ represent the width and height of the images and $i$, and $j$ correspond to the row and column indices, respectively.

The human visual system is sensitive to edges and effective preservation of these edge details can enhance the visualization of fused images. In the case of medical images, most of the textural details of the tissues are characterized by the MR image. The preservation of these crucial diagnostic edges is also important for proper demarcation among the various tissue structures. To achieve this, the gradient loss is defined as,

\begin{equation}
  L_{gradient}=\frac{1}{W\times H}\sum_{i=1}^{W}\sum_{j=1}^{H} (\triangledown F_{i,j}- \triangledown A_{i,j})^{2}
\end{equation}
where $\triangledown$ refers to the gradient operator along the horizontal and vertical direction. 

To generate a spatially consistent and artifact-free fused image, it is of prime importance that the structural similarity of the source images should be retained in the fused image. To incorporate this, a structural loss is also defined below taking into account the multi-scale structural similarity between each of the source images and the fused image. Structural loss helps to minimize the difference between the source and fused image in terms of contrast, luminescence, and structure at different scales\cite{wang2003multiscale}.

\begin{equation}
\begin{aligned}
   L_{structure}= 1-0.5*(MSSSIM(A,F)+\\
   MSSSIM(B,F))
\end{aligned}
\end{equation}

\subsection{Network architecture}
The encoder and decoder of the network follow U-Net architecture shown in  Fig.\ref{gen}. The encoder consists of four convolutional blocks having two stacked layers with one convolutional ($conv$), one Batch normalization ($Bn$), and one ReLU layer each. For each of the  $conv$ layers, the filter size is $3 \times 3$, the stride is 1 and the padding is 1. Each block in the encoder is followed by the WDEPP layer which performs edge-preserving pooling of the feature maps and downsamples the size by a factor of 2. The decoder block is symmetrical to the encoder block. It has three blocks each which consists of one transpose convolutional ($TransConv$) layer with filter size  $2 \times 2$, the stride is 2 and padding is 0 followed by an identical convolutional block as used in the encoder. The last layer is a $1 \times 1$ $conv$ layer. The skip connections are used to concatenate feature maps of the encoder with the decoder to ensure feature reuse and stabilize gradient updates during training.
\vspace {2mm}

\subsection{Training details}
For training the proposed model, the database of Whole Brain Atlas  of Harvard Medical School was used consisting of neurological MR, SPECT, CT, and PET images covering a wide variety of neoplastic,  cerebrovascular, inflammatory, infectious, and,   degenerative neurological diseases. A total of 656 image pairs consisting of 172 CT-MR pairs, 461 MR-SPECT pairs, and  23 MR-PET pairs are considered for experimentation purposes. 
The source images of size $256\times 256$ were divided into  64 x 64 size patches. Some of the patches which had little or no relevant information were discarded and finally, 4240 patches were selected and used for training. The network is trained for $30$ epochs and a batch size of 32 using Adam's optimizer with the learning rate of $1\times 10^{-3}$.  The implementation was done in the PyTorch framework, using the hardware platform with Intel Core Xeon(R) Silver 4210R CPU, 2.4 GHz, 128 GB RAM, and 24 GB NVIDIA GPU with Ubuntu 22.04 LTS 64-bit operating system.
 

\begin{figure*}[!t]
	\centering
	\scalebox{0.9}{\includegraphics[width=7.4in, 
		]{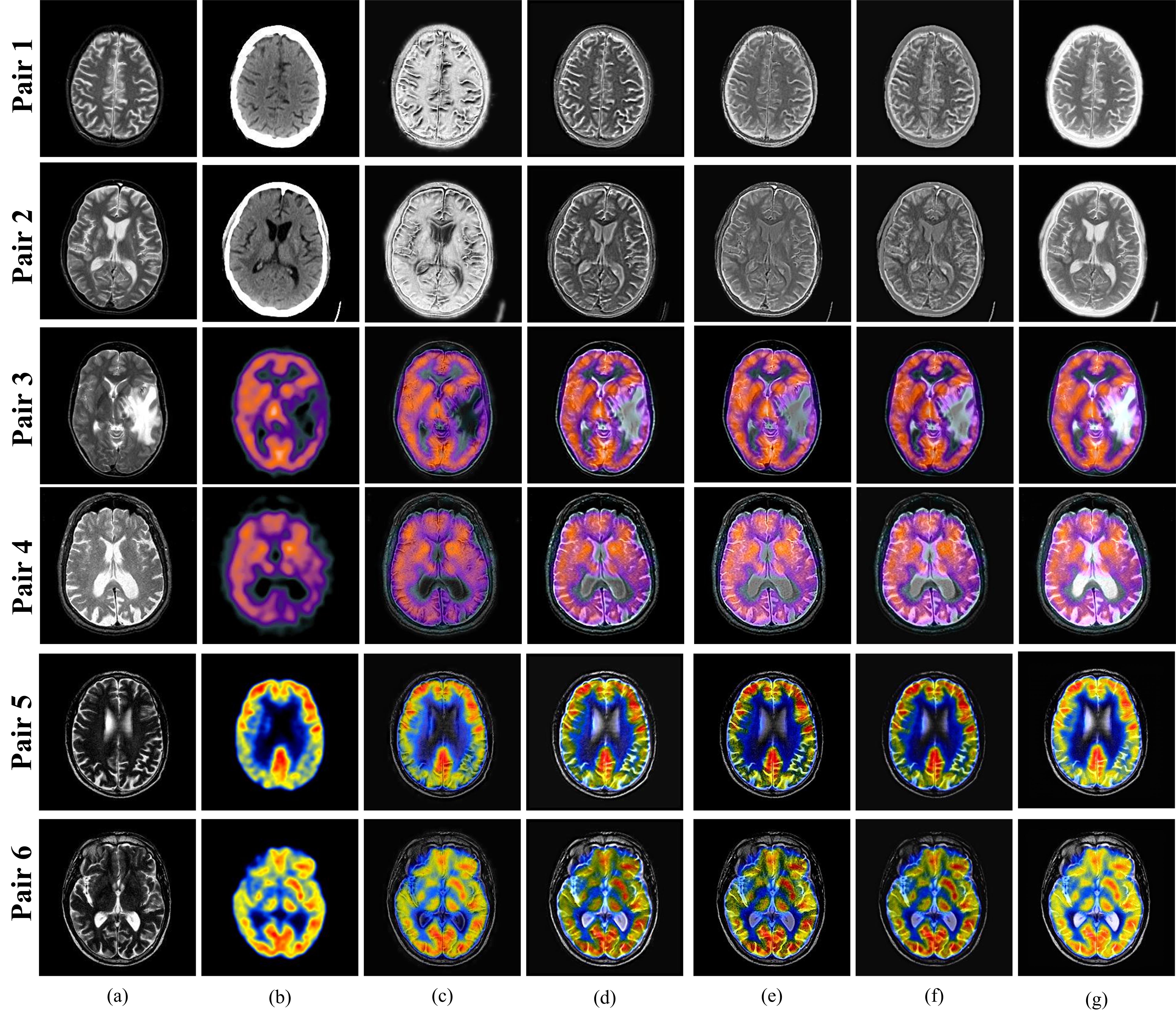}}\\
	\caption{Subjective comparison of fusion results (a) source image 1, (b) source image 2, (c) DDcGAN-2020\cite{ma2020ddcgan}, (d) DSAGAN-2021 \cite{FU2021}, (e)   SDNet-2021\cite{zhang2021sdnet}, (f) U2F-2022\cite{Xu2022}, (g) proposed method }\label{sub}
\end{figure*}

\section{Experimental Details}
To validate the performance of the proposed method, extensive experiments are carried out on the test dataset. To justify the ability of the proposed method and to fuse a variety of multimodal medical images, 100 pairs consisting of CT$-$MR-T2, SPECT, and PET images of patients suffering from various neurological disorders were considered. For performance validation, the visual and objective results are also compared with four state-of-the-art (SOTA) deep learning based fusion methods developed recently as dual-discriminator conditional GANs-based method by  Ma \textit{et al.} (DDcGAN-2020)\cite{ma2020ddcgan}, dual-stream attention mechanism (DSAGAN-2021) based method by Fu \textit{et al.} \cite{FU2021}, a squeeze-decompose network (SDNet-2021) based method by Zhang \textit{et al.} \cite{zhang2021sdnet}, a dense net based unified fusion (U2F-2022) framework by  Xu \textit{et al.} \cite{Xu2022}.
Moreover, nine different fusion metrics are considered to evaluate the proposed and existing  methods such as entropy $(EN)$, standard deviation $(SD)$  and spatial frequency $(SF)$ \cite{das2020nsst}, edge preservation $(Q_{AB/F})$ \cite{das2022tim}, mutual information $(MI)$ \cite{das2022tim}, block-wise similarity index $(Q_{C})$  \cite{das2022tim}, structural similarity index $(Q_{Y})$\cite{das2022tim}, sum of the correlations of differences $(SCD)$ \cite{aslantas2015new} and visual information fidelity for fusion  $(VIFF)$  \cite{han2013new}. Higher values of the fusion metrics signify a better fusion performance shown by the fusion approach.
 
\begin{table*}[!t]
\centering
\caption{Quantitative performance comparison of multi-modal medical image pairs shown in Fig. \ref{sub}}
\label{Table1}
\begin{tabular}{llccccccccc}
\hline
Image Pair & Method & \multicolumn{1}{l}{$EN$} & \multicolumn{1}{l}{$SD$} & \multicolumn{1}{l}{$SF$} & \multicolumn{1}{l}{$Q_{AB/F}$} & \multicolumn{1}{l}{$MI$} & \multicolumn{1}{l}{$Q_{C}$} & \multicolumn{1}{l}{$Q_{Y}$} & \multicolumn{1}{l}{$SCD$} & \multicolumn{1}{l}{$VIFF$} \\ \hline
\multirow{5}{*}{Pair 1} & DDcGAN-2020\cite{ma2020ddcgan} & {\ul 4.63} & \textbf{89.52} & \textbf{7.92} & 0.25 & 2.71 & 0.53 & 0.39 & {\ul 1.4} & 0.27 \\
 & DSAGAN-2021 \cite{FU2021} & \textbf{5.02} & 57.81 & {\ul 7.74} & 0.38 & 2.92 & \textbf{0.67} & 0.62 & 1.09 & 0.22 \\
 & SDNet-2021\cite{zhang2021sdnet} & 4.38 & 62.15 & 6.84 & 0.38 & {\ul 3.07} & 0.65 & 0.62 & 1.36 & 0.27 \\
 & U2F-2022\cite{Xu2022} & 4.31 & 58.5 & 6.86 & \textbf{0.46} & 2.98 & 0.64 & {\ul 0.63} & 1.18 & {\ul 0.35} \\
 & Proposed Method  & 4.52 & {\ul 84.73} & 6.36 & {\ul 0.43} & \textbf{3.37} & {\ul 0.66} & \textbf{0.94} & \textbf{1.53} & \textbf{0.47} \\ \hline
 \multirow{5}{*}{Pair 2} & DDcGAN-2020\cite{ma2020ddcgan} & {\ul  5.48} & \textbf{89.47} & \textbf{8.57} & 0.31 & 2.87 & 0.55 & 0.41 & {\ul 1.48} & 0.23 \\
 & DSAGAN-2021 \cite{FU2021} & \textbf{5.53} & 56.28 &  {\ul 8.22} & 0.4 & 3.03 & \textbf{0.69} & { \ul 0.65} & 1.27 & 0.18 \\
 & SDNet-2021\cite{zhang2021sdnet} & 5.00 & 56.37 & 7.36 & 0.39 & {\ul 3.14} & 0.64 & 0.61 & 1.37 & 0.18 \\
 & U2F-2022\cite{Xu2022} & 4.85 & 53.67 & 7.37 & {\ul 0.42} & 3.03 & 0.61 &  0.62 & 1.30 & {\ul 0.28} \\
 & Proposed Method  & 5.27 & {\ul 81.58} & 7.15 & \textbf{0.43} & \textbf{3.61} & {\ul 0.68} & \textbf{0.71} & \textbf{1.58} & \textbf{0.39} \\ \hline
\multirow{5}{*}{Pair 3} & DDcGAN-2020\cite{ma2020ddcgan}& {\ul 5.06} & 60.7 & 6.69 & 0.38 & 2.94 & 0.52 & 0.5 & 1.02 & 0.28 \\
 & DSAGAN-2021 \cite{FU2021} & \textbf{5.37} & {\ul 70.79} & \textbf{7.35} & 0.52 & 3.06 & 0.68 & 0.67 & 1.51 & 0.5 \\
 & SDNet-2021\cite{zhang2021sdnet} & 4.48 & 68.34 & {\ul 6.89} & 0.53 & 3.09 & \textbf{0.74} & \textbf{0.77} & \textbf{1.71} & 0.54 \\
 & U2F-2022\cite{Xu2022} & 4.78 & 64.27 & 6.75 & {\ul 0.53} & {\ul 3.21} & 0.65 & 0.67 & 1.45 & 0.57 \\
 & Proposed Method  & 4.97 & \textbf{77.62} & 6.63 & \textbf{0.6} & \textbf{3.62} & {\ul 0.71} & {\ul 0.75} & {\ul 1.68} & \textbf{0.6} \\ \hline
\multirow{5}{*}{Pair 4} & DDcGAN-2020\cite{ma2020ddcgan}&{\ul 5.73} & 59.14 & 7.79 & 0.4 & 2.92 & 0.6 & 0.55 & 0.77 & 0.23 \\
 & DSAGAN-2021 \cite{FU2021} & 5.35 & 71.44 & {\ul 8.33} & 0.56 & 3.17 & 0.78 & 0.77 & 1.21 & 0.46 \\
 & SDNet-2021\cite{zhang2021sdnet} & 5.29 & {\ul 77.12} & \textbf{8.71} & 0.6 & 3.27 & \textbf{0.81} & {\ul 0.82} & {\ul 1.6} & {\ul 0.5} \\
 & U2F-2022\cite{Xu2022} & 5.54 & 73.07 & 7.94 & {\ul 0.57} & {\ul 3.34} & 0.69 & 0.67 & 1.01 & 0.5 \\
 & Proposed Method  & \textbf{5.76} & \textbf{85.14} & 8.22 & \textbf{0.72} & \textbf{3.96} & {\ul 0.8} & \textbf{0.82} & \textbf{1.56} & \textbf{0.58} \\ \hline
\multirow{5}{*}{Pair 5} & DDcGAN-2020\cite{ma2020ddcgan} & 4.82 & 63.16 & 7.1 & 0.35 & 2.85 & 0.56 & 0.57 & 1.24 & 0.29 \\
 & DSAGAN-2021 \cite{FU2021} & \textbf{5.21} & {\ul 63.67} & {\ul 7.77} & {\ul 0.48} & 2.88 & 0.66 & 0.61 & 1.55 & 0.41 \\
 & SDNet-2021\cite{zhang2021sdnet} & 3.9 & 58.44 & \textbf{7.78} & \textbf{0.52} & {\ul 3.03} & \textbf{0.74} & {\ul 0.73} & {\ul 1.65} & 0.41 \\
 & U2F-2022\cite{Xu2022}& 4.22 & 53.72 & 6.62 & 0.42 & 3.01 & 0.61 & 0.57 & 1.47 & {\ul 0.42} \\
 & Proposed Method & {\ul 4.99} & \textbf{72.36} & 7.6 & \textbf{0.52} & \textbf{3.23} & {\ul 0.71} & \textbf{0.76} & \textbf{1.72} & \textbf{0.54} \\ \hline
\multirow{5}{*}{Pair 6} & DDcGAN-2020\cite{ma2020ddcgan} & 5.14 & {\ul 68.23} & 7.85 & 0.38 & 2.86 & 0.57 & 0.61 & 1.41 & 0.22 \\
 & DSAGAN-2021 \cite{FU2021} & \textbf{5.48} & 65.29 & 8.44 & 0.49 & 2.82 & 0.67 & 0.63 & 1.61 & 0.29 \\
 & SDNet-2021\cite{zhang2021sdnet} & 4.34 & 63.17 & \textbf{8.81} & {\ul 0.52} & {\ul 2.93} & \textbf{0.72} & {\ul 0.71} & {\ul 1.7} & {\ul 0.34} \\
 & U2F-2022\cite{Xu2022}& 4.67 & 58.02 & 7.45 & 0.41 & 2.92 & 0.6 & 0.59 & 1.57 & 0.33 \\
 & Proposed Method & {\ul 5.42} & \textbf{80.17} & {\ul 8.47} & \textbf{0.55} & \textbf{3.27} & {\ul 0.7} & \textbf{0.77} & \textbf{1.75} & \textbf{0.38} \\ \hline
\end{tabular}
\end{table*}

\begin{table*}[!t]
\scriptsize
\centering
\caption{Averaged performance analysis of fusion methods for 100 pairs of multi-modal medical images\\(Average $\pm$ Standard deviation (Score)) }
\label{Table2}
\begin{tabular}{lcclllll}
\hline
\begin{tabular}[c]{@{}l@{}}Performance \\ Metric\end{tabular} & \multicolumn{1}{l}{MR Image} & \multicolumn{1}{l}{\begin{tabular}[c]{@{}l@{}}CT/SPECT/\\ PET Image\end{tabular}} & DDcGAN-2020\cite{ma2020ddcgan} & DSAGAN-2021 \cite{FU2021} & SDNet-2021\cite{zhang2021sdnet}& U2F-2022\cite{Xu2022} & Proposed  method \\ \hline
$EN$ & \multicolumn{1}{l}{4.49  ± 0.65} & \multicolumn{1}{l}{3.8  ± 0.85} & 5.4  ± 0.7 (4) & 5.41  ± 0.39 (5) & 4.68  ± 0.71 (1) & 4.74  ± 0.58 (2) & 5.22  ± 0.64 (3) \\
$SD$ & \multicolumn{1}{l}{57.63  ± 7.4} & \multicolumn{1}{l}{71.49  ± 16.32} & 74.14  ± 16.71 (4) & 60.95  ± 5.48 (2) & 61.7  ± 6.48 (3) & 56.87  ± 7.12 (1) & 77.52  ± 9.3 (5) \\
$SF$ & \multicolumn{1}{l}{7.23  ± 0.74} & \multicolumn{1}{l}{5.2  ± 0.9} & 7.81  ± 0.73 (4) & 7.9  ± 0.54 (5) & 7.45  ± 0.71 (3) & 6.97  ± 0.59 (1) & 7.21 ± 0.72 (2) \\
$Q_{AB/F}$ & - & - & 0.37  ± 0.08 (1) & 0.45  ± 0.09 (2) & 0.5  ± 0.09 (4) & 0.47  ± 0.05 (3) & 0.51  ± 0.14 (5) \\
$MI$ & - & - & 2.73  ± 0.25 (1) & 2.85  ± 0.27 (2) & 3.05  ± 0.3 (4) & 2.97  ± 0.29 (3) & 3.43  ± 0.35 (5) \\
$Q_{C}$ & - & - & 0.55  ± 0.04 (1) & 0.66  ± 0.05 (3) & 0.69  ± 0.08 (4) & 0.59  ± 0.06 (2) & 0.69 ± 0.07 (5) \\
$Q_{Y}$ & - & - & 0.49  ± 0.07 (1) & 0.63  ± 0.06 (3) & 0.69  ± 0.1 (4) & 0.59  ± 0.06 (2) & 0.72 ± 0.06 (5) \\
$SCD$ & - & - & 1.27  ± 0.32 (2) & 1.3  ± 0.27 (3) & 1.48  ± 0.21 (4) & 1.27  ± 0.25 (1) & 1.58  ± 0.14 (5) \\
$VIFF$ & - & - & 0.27  ± 0.05 (1) & 0.33  ± 0.17 (2) & 0.37  ± 0.14 (3) & 0.39  ± 0.11 (4) & 0.48  ± 0.1(5) \\
\begin{tabular}[c]{@{}l@{}}Average \\ Score\end{tabular} & - & - & 2.11 & 3 & 3.33 & 2.11 & 4.44 \\ 
Run time (Sec)& - & - & 1.44  ± 0.097 & 0.009 ± 0.001  & 0.033  ± 0.004 & 0.659  ± 0.063 & 0.018  ± 0.002 \\\hline
\end{tabular}
\end{table*}

\section{Results and Discussion} 
\subsection{Visual analysis}
Fig. \ref{sub} shows six image pairs along with the fused images obtained by the four SOTA fusion methods and the proposed method. For MR-T2$-$CT fusion (pair 1 and 2), the DDcGAN-2020 \cite{ma2020ddcgan} method gives fused images with visible artifacts resulting in poor  visual results (refer Fig. \ref{sub}(c)). The  DSAGAN-2021 \cite{FU2021}, SDnet-2021 \cite{zhang2021sdnet} and U2F-2022 \cite{Xu2022} methods offer a better representation of soft tissue details of MR-T2 image, however, lose the hard tissue information present in the CT image, as a result, the skull boundary is not highlighted well which can be visualized from Figs. \ref{sub}(d)-(f). On the other hand, Fig. \ref{sub}(g)  shows that the proposed method yields a fused image with effective preservation of intensity and textures of both hard and soft tissues. For MR-T2$-$SPECT image fusion (pair 3 and 4) depicting a case of metastatic bronchogenic carcinoma, DDcGAN-2020 \cite{ma2020ddcgan} fusion method ceases to preserve the textural information of the MR-T2 image. The SDNet-2021 \cite{zhang2021sdnet} and U2F-2022 \cite{Xu2022} methods capture the gradients of the soft tissues, however these methods are not able to preserve the color maps of the SPECT image, leading to reduced contrast and visually inferior fused images. Though the  DSAGAN-2021 \cite{FU2021} method and the proposed method offer better contrast but the tumor is highlighted and demarcated better in the fused images obtained by the proposed method (refer to Fig. \ref{sub}(a), (d) and (f)). For MR-T2$-$PET image fusion (pair 5 and 6), it can be visualized from Fig. \ref{sub}, that the proposed method offers better integration and retention of both the anatomical and functional details of the tissues in terms of contrast, edge preservation, spatial and color fidelity as compared to other SOTA fusion methods. \\

\subsection{Quantitative analysis}
The quantitative performance metrics of the fused images shown in Figs. \ref{sub} are mentioned in Table  \ref{Table1}. The highest-performing method is presented in bold and the second-highest is underlined. It can be observed that the proposed method achieves higher values of most of the metrics as compared to other fusion methods. It ranks first for $MI$ and $VIFF$ metrics indicating a higher extent of information content and visual quality of the fused image.  Furthermore, to demonstrate a concise analysis of the overall fusion performance, all performance evaluation metrics are evaluated for 100 test pairs and their average and the standard deviation are tabulated in Table \ref{Table2}. Each method is also given a score between 1 to 5. The lowest-performing method is given a score of 1, while a score of 5 is given to the highest-performing one. Table \ref{Table2} indicates that the proposed method gets the highest score for metrics $SD$, $Q_{AB/F}$, $MI$, $Q_{C}$, $Q_{Y}$, $SCD$ and $VIFF$ and also achieves an average overall score which validates subjective results presented in the previous section. The DSAGAN-2021 \cite{FU2021} method ranks first and gets slightly higher values $SF$ compared to the proposed method but lags in achieving similar performance for other metrics especially $SD$ and $VIFF$.The average run time of all the methods for fusion of one image pair of size $256\times256$  is also shown in Table \ref{Table2}. The proposed method takes about 0.0018 seconds which is much lesser than the time taken by most of the other methods. Following is the summary of the average performance gain achieved by the proposed method over other SOTA methods;

\begin{enumerate}
     \item  The proposed method gets 4.55\%- 36.32\%  and 23.33\%- 78.14\%  higher values of $SD$  and $VIFF$, respectively indicating higher visual fidelity of fused images with better contrast and color consistency.
    \item It achieves 3.62\%- 38.17\% higher values of $Q_{AB/F}$ indicating higher preservation of fine tissue edges offering better boundary preservation and improved demarcation among tissue structures.
    \item It provides 12.72\%- 25.82\% higher $MI$ values referring to higher preservation of characteristic information of the source images. 
    \item  It gets 9.52\%- 25.45\%  and 4.35\%- 46.94\%  higher values of $Q_{C}$ and $Q_{Y}$ respectively indicating higher visual structural similarity between the source and fused images.
	\item It achieves 6.76\% - 24.40\% higher values of $SCD$ indicating a higher Correlation among the fused and source images.
\end{enumerate}

\begin{figure}[!t]
	\centering
	\scalebox{0.45}{\includegraphics[width=7.4in, 
		]{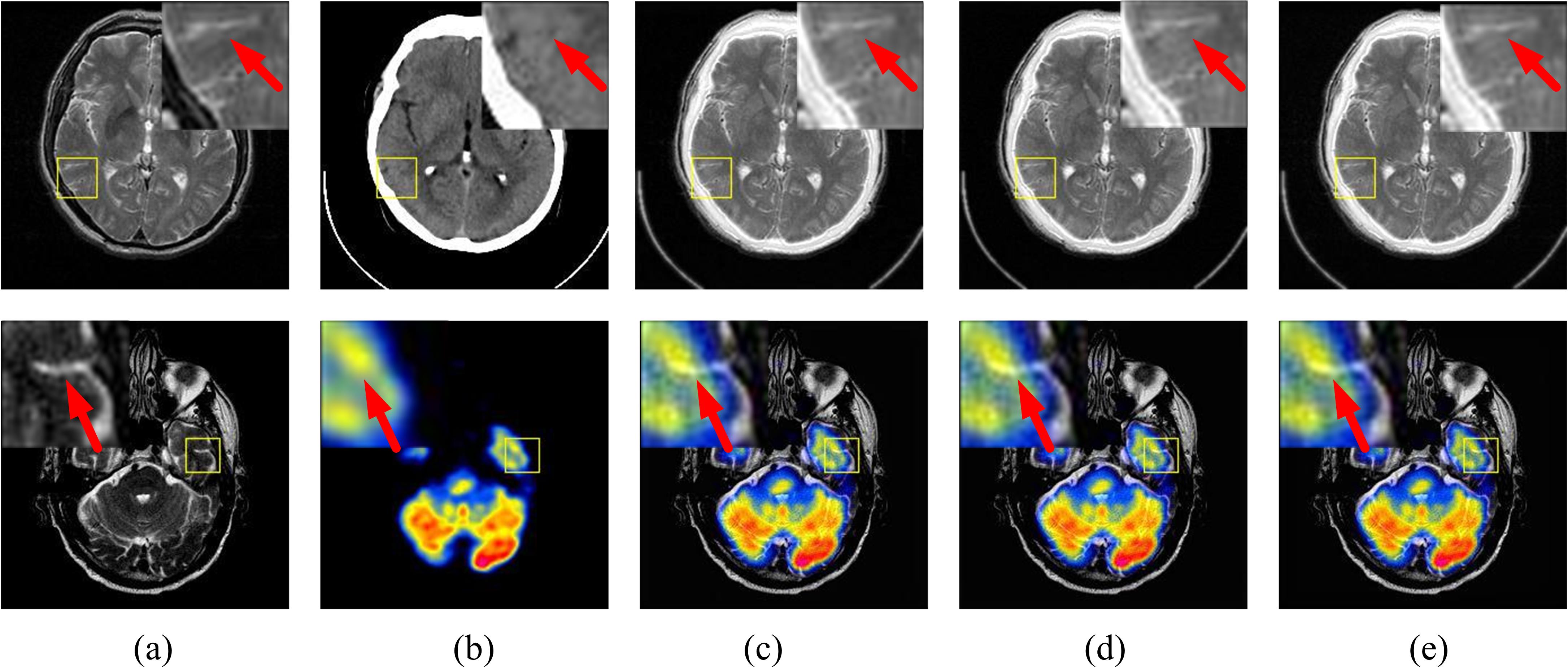}}\\
	\caption{Subjective comparison of fusion results (a) MR T2 image, (b) CT/PET image,  results obtained by using(c) max pooling, (d) average pooling, (e)  WDEPP }\label{abl1}
\end{figure}
\subsection{Ablation study}
This section validates the effect of using WDEPP in place of conventional max pooling and average pooling on the fusion performance of the proposed method. For such purpose, the WDEPP layer is replaced with a max/average pooling layer of filter size $2$ and stride $1$. The rest of the layers in the encoder and decoder remains unchanged as shown in Fig. \ref{gen}. Fig. \ref{abl1} shows source image pairs of MR-T2$-$CT, MR-T2$-$SPECT and the corresponding fusion results obtained by using various pooling strategies. Their objective results are also tabulated in Table \ref{ablt}. From Fig. \ref{abl1}, it can be visualized that 
WDEPP approach preserves the fine edges of the MR image with better contrast and clarity (refer to the zoomed regions in Fig. \ref{abl1}). Higher values of $Q_{AB/F}$ and $SF$ presented in Table \ref{ablt} also validate the visual results shown with the WDEPP approach. Moreover, the WDEPP approach also gets 11.81\%, 13.68\%, and   3.07\%, 5.91\%  higher values $MI$ for MR-T2$-$CT and MR-T2$-$SPECT image fusion, respectively, referring to its improved ability to extract and fuse the complementary information of the source images.

Furthermore, the average analysis is also carried out and presented in Table \ref{ablt_avg}. It can be inferred from the results that the WDEPP approach gets higher values of most of the fusion metrics compared to the conventional max and average pool approaches.  Therefore, it can be concluded that replacing the conventional pooling layers with the WDEPP layer in the proposed autoencoder architecture helps to extract more finer and pertinent features from the source images and improves the overall fusion performance.

\begin{table}[!t]
\scriptsize
\centering
\caption{Quantitative performance comparison of  pooling strategies for fused image shown in Fig. \ref{abl1} }
\label{ablt}
\begin{tabular}{
>{\columncolor[HTML]{FFFFFF}}l 
>{\columncolor[HTML]{FFFFFF}}c 
>{\columncolor[HTML]{FFFFFF}}c 
>{\columncolor[HTML]{FFFFFF}}c 
>{\columncolor[HTML]{FFFFFF}}c 
>{\columncolor[HTML]{FFFFFF}}c 
>{\columncolor[HTML]{FFFFFF}}c }
\hline
\cellcolor[HTML]{FFFFFF} & \multicolumn{3}{c}{\cellcolor[HTML]{FFFFFF}Pair 1} & \multicolumn{3}{c}{\cellcolor[HTML]{FFFFFF}Pair 2} \\ \cline{2-7} 
\multirow{-2}{*}{\cellcolor[HTML]{FFFFFF}\begin{tabular}[c]{@{}c@{}}Performance\\   Metric\end{tabular}} & \begin{tabular}[c]{@{}c@{}}Max \\ Pool\end{tabular} & \begin{tabular}[c]{@{}c@{}}Average \\ Pool\end{tabular} & WDEPP & \begin{tabular}[c]{@{}c@{}}Max \\ Pool\end{tabular} & \begin{tabular}[c]{@{}c@{}}Average\\  Pool\end{tabular} & WDEPP \\ \hline
$EN$ & 6.173 & 6.094 & \textbf{6.35} & 5.063 & 4.866 & \textbf{5.167} \\
$SD$ & \textbf{75.114} & 74.363 & 74.505 & 64.199 & 64.145 & \textbf{64.304} \\
$SF$ & 7.168 & 7.179 & \textbf{7.242} & 8.232 & 8.093 & \textbf{8.312} \\
$Q_{AB/F}$ & 0.376 & 0.377 & \textbf{0.396}  & 0.733 & 0.721 & \textbf{0.74} \\
$MI$ & 3.717 & 3.656 & \textbf{4.156} & 2.833 & 2.757 & \textbf{2.92} \\
$Q_{C}$ & 0.344 & 0.262 & \textbf{0.703} & 0.756 & 0.729 & \textbf{0.796} \\
$Q_{Y}$ & 0.505 & 0.464 & \textbf{0.739} & 0.8 & 0.773 & \textbf{0.824} \\
$SCD$ & 1.407 & \textbf{1.416} & 1.402 & 1.877 & \textbf{1.881} & 1.872 \\
$VIFF$ & 0.376 & \textbf{0.379} & 0.37 & 0.529 & \textbf{0.53} & 0.526 \\ \hline
\end{tabular}
\end{table}

\begin{table}[!t]
\scriptsize
\centering
\caption{Averaged fusion performance of the proposed method for different Pooling strategies \\(Average $\pm$ Standard deviation) }
\label{ablt_avg}
\begin{tabular}{llll}
\hline
\multirow{2}{*}{\begin{tabular}[c]{@{}l@{}}Performance   \\ Metrics\end{tabular}} & \multicolumn{3}{c}{Pooling Approach} \\ \cline{2-4} 
 & Max pooling & Average pooling & WDEPP\\ \hline
$EN$ & \textbf{5.36 ± 0.56} & 5.16 ± 0.57 & 5.22 ± 0.64 \\
$SD$ & 77.26 ± 9.47 & 77.28 ± 9.63 & \textbf{77.52 ± 9.3} \\
$SF$ & 7.17 ± 0.72 & 7.14 ± 0.69 & \textbf{7.21 ± 0.72} \\
$Q_{AB/F}$ & 0.5 ± 0.14 & 0.49 ± 0.13  & \textbf{0.51 ± 0.14} \\
$MI$ & 3.32 ± 0.3 & 3.26 ± 0.29     & \textbf{3.43 ± 0.35} \\
$Q_{C}$ & 0.63 ± 0.1 & 0.61 ± 0.11 & \textbf{0.69 ± 0.07} \\
$Q_{Y}$ & 0.67 ± 0.08 & 0.66 ± 0.08 & \textbf{0.72 ± 0.06} \\
$SCD$ & 1.57 ± 0.14 & \textbf{1.59 ± 0.13} & 1.58 ± 0.14 \\
$VIFF$ & 0.47 ± 0.09 & \textbf{0.48 ± 0.09} & \textbf{0.48 ± 0.1} \\
Run time (Sec) & \textbf{0.008 ± 0.001} & 0.009 ± 0.001 & 0.018 ± 0.002 \\
\hline
\end{tabular}
\end{table}


\section{Conclusions} \label{Conclusion}
This paper presents an edge-preserving autoencoder-based multi-modal neurological image fusion framework. In the proposed approach, the conventional pooling layers of the encoder have been replaced by the WDEPP layer which reduces the size of feature maps while retaining the fine edges and textures intact in the fused images. The wavelet decomposition of the feature maps and individual channel attention to the decomposed wavelet sub-bands helps to fuse the global and local information of the source images effectively which leads to fused images with better visual contrast and textural clarity. The visual and quantitative performance also justifies the efficacy of the proposed method by fusing a variety of neurological image pairs. The proposed method also outperforms the existing fusion methods and demonstrates a notable improvement in edge and information preservation with higher visual contrast and clarity of the fused images. The proposed method is also efficient in terms  of the time consumption and hence can assist radiologists in more efficient, faster, and reliable fusion for improved diagnosis and treatment.
\ifCLASSOPTIONcaptionsoff
  \newpage
\fi
\bibliographystyle{ieeetr} 
\bibliography{Main}

\begin{thebibliography}{10}

\bibitem{li2017pixel}
S.~Li, X.~Kang, L.~Fang, J.~Hu, and H.~Yin, ``Pixel-level image fusion: A
  survey of the state of the art,'' {\em Information Fusion}, vol.~33,
  pp.~100--112, 2017.

\bibitem{HERMESSI2021108036}
H.~Hermessi, O.~Mourali, and E.~Zagrouba, ``Multimodal medical image fusion
  review: Theoretical background and recent advances,'' {\em Signal
  Processing}, vol.~183, p.~108036, 2021.

\bibitem{dogra2017multi}
A.~Dogra, B.~Goyal, and S.~Agrawal, ``From multi-scale decomposition to
  non-multi-scale decomposition methods: a comprehensive survey of image fusion
  techniques and its applications,'' {\em IEEE Access}, vol.~5,
  pp.~16040--16067, 2017.

\bibitem{du2016overview}
J.~Du, W.~Li, K.~Lu, and B.~Xiao, ``An overview of multi-modal medical image
  fusion,'' {\em Neurocomputing}, vol.~215, pp.~3--20, 2016.

\bibitem{liu2017medical}
Y.~Liu, X.~Chen, J.~Cheng, and H.~Peng, ``A medical image fusion method based
  on convolutional neural networks,'' in {\em 20th International Conference on
  Information Fusion}, pp.~1--7, 2017.

\bibitem{liu2016image}
Y.~Liu, X.~Chen, R.~K. Ward, and Z.~J. Wang, ``Image fusion with convolutional
  sparse representation,'' {\em IEEE Signal Processing Letters}, vol.~23,
  no.~12, pp.~1882--1886, 2016.

\bibitem{li2018densefuse}
H.~Li and X.-J. Wu, ``{DenseFuse: A fusion approach to infrared and visible
  images},'' {\em IEEE Transactions on Image Processing}, vol.~28, no.~5,
  pp.~2614--2623, 2018.

\bibitem{Xu2022}
H.~Xu, J.~Ma, J.~Jiang, X.~Guo, and H.~Ling, ``U2fusion: A unified unsupervised
  image fusion network,'' {\em IEEE Transactions on Pattern Analysis and
  Machine Intelligence}, vol.~44, no.~1, pp.~502--518, 2022.

\bibitem{zhang2021sdnet}
H.~Zhang and J.~Ma, ``{SDNet: A versatile squeeze-and-decomposition network for
  real-time image fusion},'' {\em International Journal of Computer Vision},
  vol.~129, no.~10, pp.~2761--2785, 2021.

\bibitem{ma2020ddcgan}
J.~Ma, H.~Xu, J.~Jiang, X.~Mei, and X.-P. Zhang, ``{DDcGAN: A
  dual-discriminator conditional generative adversarial network for
  multi-resolution image fusion},'' {\em IEEE Transactions on Image
  Processing}, vol.~29, pp.~4980--4995, 2020.

\bibitem{FU2021}
J.~Fu, W.~Li, J.~Du, and L.~Xu, ``{DSAGAN: A generative adversarial network
  based on dual-stream attention mechanism for anatomical and functional image
  fusion},'' {\em Information Sciences}, vol.~576, pp.~484--506, 2021.

\bibitem{xu2021emfusion}
H.~Xu and J.~Ma, ``{EMFusion: An unsupervised enhanced medical image fusion
  network},'' {\em Information Fusion}, vol.~76, pp.~177--186, 2021.

\bibitem{fan1906semantic}
F.~Fan, Y.~Huang, L.~Wang, X.~Xiong, Z.~Jiang, Z.~Zhang, and J.~Zhan, ``A
  semantic-based medical image fusion,'' {\em arXiv preprint arXiv:1906.00225}.

\bibitem{sineesh2021exploring}
A.~Sineesh and M.~R. Panicker, ``Exploring novel pooling strategies for edge
  preserved feature maps in convolutional neural networks,'' {\em arXiv
  preprint arXiv:2110.08842}, 2021.

\bibitem{li2020wavelet}
Q.~Li, L.~Shen, S.~Guo, and Z.~Lai, ``{Wavelet integrated CNNs for noise-robust
  image classification},'' in {\em Proceedings of the IEEE/CVF Conference on
  Computer Vision and Pattern Recognition}, pp.~7245--7254, 2020.

\bibitem{das2022tim}
M.~Das, D.~Gupta, P.~Radeva, and A.~M. Bakde, ``Optimized multimodal
  neurological image fusion based on low-rank texture prior decomposition and
  super-pixel segmentation,'' {\em IEEE Transactions on Instrumentation and
  Measurement}, pp.~1--9, 2022.

\bibitem{das2020nsst}
M.~Das, D.~Gupta, P.~Radeva, and A.~M. Bakde, ``Nsst domain ct--mr neurological
  image fusion using optimised biologically inspired neural network,'' {\em IET
  Image Processing}, vol.~14, no.~16, pp.~4291--4305, 2020.

\bibitem{wang2003multiscale}
Z.~Wang, E.~P. Simoncelli, and A.~C. Bovik, ``Multiscale structural similarity
  for image quality assessment,'' in {\em The Thrity-Seventh Asilomar
  Conference on Signals, Systems \& Computers}, vol.~2, pp.~1398--1402, 2003.

\bibitem{aslantas2015new}
V.~Aslantas and E.~Bendes, ``A new image quality metric for image fusion: the
  sum of the correlations of differences,'' {\em AEU-International Journal of
  Electronics and Communications}, vol.~69, no.~12, pp.~1890--1896, 2015.

\bibitem{han2013new}
Y.~Han, Y.~Cai, Y.~Cao, and X.~Xu, ``A new image fusion performance metric
  based on visual information fidelity,'' {\em Information Fusion}, vol.~14,
  no.~2, pp.~127--135, 2013.

\end{thebibliography}

\vspace{-10mm}

\end{document}